\documentclass{article}
\usepackage[utf8]{inputenc}
\usepackage{authblk}
\usepackage{hyperref}
\usepackage{bm}
\usepackage{xcolor}
\usepackage{graphicx}
\usepackage{setspace}
\usepackage{braket}
\usepackage{amsmath}
\usepackage{amssymb}
\usepackage{physics}
\usepackage{mathtools}

\title {Reframing the Event Horizon: The Harlow-Hayden Computational Approach to the Firewall Paradox}
\author{Galina Weinstein}
\affil{\normalsize Reichman University, The Efi Arazi School of Computer Science, Herzliya; University of Haifa, The Department of Philosophy, Haifa, Israel.} 

\begin{document}

\maketitle

\begin{abstract}
This study critically reevaluates the Harlow-Hayden (HH) solution to the black hole information paradox and its articulation in the firewall paradox. The exploration recognizes the HH solution as a revolutionary approach in black hole physics, steering away from traditional constraints to depict the event horizon as a computational rather than a physical barrier. The paper first maps the initial physical dilemma that instigated the HH journey, introducing Alice, an observer facing intricate computational challenges as she approaches the black hole. I then depict the evolution of the narrative, describing how Alice was facilitated with a quantum computer to surmount the computational challenges and further detailing the augmented complexities arising from the integration of the physical dynamics of the black hole. Yet, HH's research applies the AdS/CFT correspondence to explore the dynamic unitary transformation in solving the firewall paradox through decoding Hawking radiation. However, it identifies a contradiction; the eternal perspective of black holes from the AdS/CFT theory challenges the firewall paradox's foundation. Finally, I narrate a paradigm shift as HH reframes Alice’s task within the realms of error-correcting codes, illustrating a remarkable transition from a physical problem in black hole physics to a computational predicament in computer science. The study revisits pivotal moments in understanding black hole physics ten years later through this reexamination. 
\end{abstract}

\section{Introduction}

In this study, I re-examine the Harlow-Hayden (HH) solution \cite{Harlow} to the black hole information paradox and its manifestation in the firewall paradox \cite{AMPS}. The initiation of the HH solution heralded a pivotal moment in studying black hole physics. It deviated from the traditional analysis confined by spatial and temporal borders, proposing a fresh perspective wherein the event horizon is perceived more as a barrier dictated by computational hurdles rather than mere physical limits.

I initiate in section \ref{1} with a concise historical introduction where I elucidate the concepts of Hawking radiation, the monogamy of entanglement, and the firewall paradox, setting a firm foundation for the ensuing discussion. Following this, I delve into a detailed re-evaluation of the HH solution to the firewall paradox. 

The initial approach of HH was grounded in addressing a physical dilemma, a prevalent strategy to untangle the intricacies of the Hawking information paradox during that period, explored in detail in section \ref{2}. Alice, the observer descending into the black hole, faces a computationally daunting challenge. To navigate this, HH empowered Alice with a quantum computer; a development unfolded in section \ref{3}. The narrative evolves by integrating physical insights concerning black hole dynamics in section \ref{4}, yet Alice encounters additional complexities.

Adapting to the unfolding scenario, HH redefined Alice's mission in the context of error-correcting codes, elucidated in section \ref{5}. Epistemically, this transformative journey took us from grappling with a physical conundrum rooted in black hole physics to ultimately contending with a computational issue in the domain of computer science.

\section{Hawking radiation and the firewall paradox} \label{1}

According to classical general relativity, nothing can escape from the event horizon of a black hole. However, in 1974, Hawking penned a letter to \emph{Nature}, provocatively titled "A Black Hole Explosion?". In this letter, Hawking intended to demonstrate that significant quantum effects might be associated with black holes \cite{Hawking1}. In a reflection three years later, Hawking detailed the initial realization that catalyzed his novel hypothesis, stating, "To my great surprise, I found," back in 1974, "that the black hole seemed to emit particles at a steady rate. Like everyone else at that time, I accepted the dictum that a black hole could not emit anything. I therefore invested substantial effort into dismissing this unsettling result, but it persistently refused to disappear, forcing me to eventually accept it" \cite{Hawking3}.  

Crucially, this hypothetical radiation arises due to how observers at infinity categorize scalar field modes. This classification is discontinuous at the black hole's horizon and disregards all information about the modes within the horizon. Contrarily, an observer plunging into the black hole wouldn't perceive any particle creation, as they wouldn't employ such a discontinuous division but rather analyze the field through modes continuous at the event horizon \cite{Hawking2}.

Hawking radiation can be conceptualized as follows: Near the event horizon of a black hole, a particle-antiparticle pair spontaneously forms due to quantum fluctuations. One particle, which we will denote as $A$, falls into the black hole possessing negative energy, while the other particle, $B$, escapes with positive energy. Despite being separated, particles $A$ and $B$ remain entangled, ensuring energy conservation in the system. Over time, this process leads to continuous emission of particles ($B$) away from the black hole, a phenomenon termed Hawking radiation. Meanwhile, the absorption of negative energy particles ($A$) gradually decreases the black hole's mass and energy, eventually leading to its complete evaporation.

We face a perplexing question when we consider what happens to the information that enters a black hole. According to the no-hair theorem, a black hole is characterized solely by three parameters: mass, angular momentum, and electric charge. It ostensibly retains no other details about the matter it engulfs, suggesting that a vast amount of information becomes irrevocably lost during the process of gravitational collapse. This proposition seems to violate the second law of thermodynamics, which would imply that the black hole should have zero entropy as all the information (with high entropy) is eradicated.

However, this theory confronts a significant challenge from quantum mechanics, specifically the principles articulated in quantum field theory and the principle of unitarity. The latter insists on the conservation of information, which contests the assertion that information entering a black hole is irretrievably lost. Moreover, if information is preserved, the black hole would have non-zero entropy, giving it a finite temperature and leading it to emit thermal Hawking radiation. This process would theoretically result in the black hole gradually losing mass until it evaporates entirely, leaving only Hawking radiation composed of certain particles behind. This conservation of information gives rise to what is known as the black hole information paradox, a fundamental conflict between quantum mechanics and the theory of general relativity. This paradox underlines a critical discrepancy in our understanding of modern physics: if the information is indeed lost in a black hole, it undermines the foundations of quantum mechanics; conversely, if the information is conserved, it challenges the principles of classical general relativity.

In 1993, Leonard Susskind, Lárus Thorlacius, and John Uglum introduced a solution to the black hole information paradox known as "black hole complementarity." This principle hinges on the experiences of two observers: Alice, who falls into the black hole, and Bob, who remains outside of it. According to this proposal, the conflicting descriptions of the black hole interior, as observed by Alice and Bob, are not contradictory but rather complementary, hence the term "black hole complementarity" \cite{Susskind1}; \cite{Susskind2}.

From Bob's perspective, observing from a safe distance outside the event horizon, the horizon acts as a physical membrane, becoming a hot layer just above the black hole's horizon, termed the "stretched horizon." In this perspective, Alice appears to become increasingly red-shifted as she approaches the event horizon, effectively getting "frozen" at the horizon. Alice would never seem to cross the event horizon; instead, she would get incinerated due to the extremely high temperatures. The black hole then re-emits her mass energy in the form of Hawking radiation, which carries the information about Alice, allowing Bob to theoretically reconstruct her from the information contained in the radiation, thereby preserving the principle of unitarity in quantum mechanics.

Conversely, from Alice’s vantage point, as she falls towards the black hole, due to the equivalence principle, she doesn't notice anything unusual at the moment she crosses the event horizon, experiencing a "no drama" scenario. Contrary to Bob's observations, she wouldn't see herself getting stuck at the horizon or getting incinerated but would smoothly pass the event horizon and eventually meet her fate at the singularity at the black hole’s core, seemingly violating the principle of unitarity due to the apparent loss of information as she crosses the event horizon.

Black hole complementarity proposes a dual reality where Alice's and Bob’s descriptions are correct in their respective frames of reference. However, they cannot communicate and compare notes after Alice crosses the horizon. This approach offers a potential resolution to the black hole information paradox, albeit at the cost of introducing a vexing puzzle tied to the nature of entanglement and the very principles of quantum mechanics.

In this paradox, we envisage three particles: $A$, $B$, and $C$, where $A$ and $B$ are a pair of entangled particles and particle $C$ shares information with $B$. Particle $A$ is swallowed by the black hole, $B$ is emitted as Hawking radiation, and particle $C$ is another piece of radiation emitted before $B$, creating a mixed state of $A$, $B$, and $C$. Imagine a scenario involving Alice, an observer who first measures early radiation from particle $C$, then does the same for the later radiation from particle $B$ before crossing the event horizon to encounter particle $A$, which shares entanglement with particle $B$, and through it, with particle $C$. This scenario raises critical questions grounded in the concept of monogamy of entanglement (the principle that a quantum system can only be entangled with one system at a time), hinting at a violation of the no-cloning theorem of quantum mechanics. Having measured both $B$ and $C$, Alice carries this information into the black hole where she measures $A$, suggesting the duplication of information, a direct contradiction to the no-cloning theorem prohibiting the exact replication of arbitrary unknown quantum states.

Seeking to resolve this deep-seated paradox, Ahmed Almheiri, Donald Marolf, Joseph Polchinski, and James Sully (AMPS) reformulated the black hole information paradox, presenting a new paradox, the “firewall paradox.” AMPS brought the information paradox to a more concrete footing by arguing that if the information about the in-falling matter is to be preserved and able to be retrieved from a black hole, as suggested by unitary evolution in quantum mechanics, then there must be a highly energetic Planck-scale firewall at the event horizon, which would effectively destroy anything falling in, thus preserving the information. This conclusion, however, starkly contradicts the general relativistic prediction of a smooth event horizon, as well as the equivalence principle (“absence of drama for the infalling observer”), which states that falling through a horizon should be uneventful for an in-falling observer \cite{AMPS}.

In 2013, Daniel Harlow and Patrick Hayden (HH) proposed an innovative approach to address the firewall paradox's contentious issues. In this approach, they suggested that Alice could theoretically verify the transmission of quantum information emanating from the interior of a black hole, but to do this successfully, she would require a quantum computer equipped with an error-correcting code potent enough to handle computations of extraordinary complexity \cite{Harlow}. 

In the subsequent discussion, I will demonstrate that, from an epistemic standpoint, the HH paper fundamentally altered the framework and tenor of the existing discourse and forged a novel pathway for engagement with the firewall paradox. Leveraging insights from computational theory unveiled new pathways for understanding and possibly resolving the firewall paradox.

\section{HH begin with a physical problem} \label{2}

\subsection{Monogamy of entanglement and the Firewall paradox}

HH consider the quantum description of a Schwarzschild black hole in 3 + 1 dimensions from the points of view of an external observer, Charlie, and an infalling observer, Alice. The entropy of the black hole is proportional to $M^2$ in Planck units, where $M$ is the mass of the black hole. The time it takes for the black hole to evaporate is proportional to $M^3$. Alice has a task at hand to extract information from around $n\sim M^2$ bits of Hawking radiation within a time frame $T\sim n^{3/2}$ before the black hole evaporates completely. Alice needs to apply a unitary transformation to the Hawking radiation to extract the desired information. This process effectively unscrambles the desired information, making it accessible in a specific subfactor of the Hilbert space. Charlie is positioned far from the black hole, at infinity. This position gives him ample time and memory resources to measure the Hawking radiation emitted by the black hole with great precision. 
\vspace{1mm} 

\emph{Let us start by considering Charlie's perspective}. Charlie's description is based on the following three postulates \cite{Harlow}:

1) Charlie postulates that the black hole's formation and subsequent evaporation can be described as a unitary process. 

2) Furthermore, we can conceptualize the system as undergoing either continuous or discrete time evolution, wherein at any individual moment, it is in a pure quantum state $\ket{\psi}$, which lives in a specified Hilbert space labeled $H_{\text{outside}}$. Charlie breaks down $H_{\text{outside}}$ into three subspaces: 
$H_{\text{outside}}= H_H \otimes H_B \otimes H_R$, 
where $H_H$ represents the degrees of freedom inside or close to the black hole. Heuristically associated with the "stretched horizon" at a Schwarzschild coordinate radius given by $r=2GM+\epsilon$, $G$ is the gravitational constant, $M$ is the black hole's mass, and $\epsilon$ is some ultraviolet cutoff. 
$H_B$ represents the field theory modes in the near-horizon region of the black hole. This region has a Schwarzschild coordinate radius in the range $2GM+\epsilon<r<3GM$, indicating that it is close to the black hole but outside the stretched horizon described in $H_H$. The geometry in this region is close to that of Rindler space. It will include modes with Schwarzschild energy less than the black hole temperature $T = \frac{1}{4\pi GM}$. Modes with higher energy are not confined to this near-horizon region and are considered part of $H_R$. $H_R$ represents the Hawking radiation field, the modes of the radiation field outside the black hole, with Schwarzschild coordinate radius $r>3GM$. $H_R$ includes higher energy modes that are not confined to the near-horizon region described in $H_B$. 

We can relabel the different regions associated with a black hole and the surrounding environment as \(B\), \(R\), and \(H\). 
\(B\) is the zone just outside the event horizon of the black hole. \(B\) is where one can theoretically observe phenomena related to the near-horizon dynamics of the black hole, including the effects of Hawking radiation and other quantum gravitational effects. Subscript \(B\) ties it to phenomena and quantum information related to \(B\). 
\(H\) refers to the black hole's horizon, and \(R\) represents the radiation field outside the black hole, in a region farther from the black hole, where one can observe the radiation emitted by the black hole, including the Hawking radiation that has escaped the gravitational pull of the black hole. Subscript \(R\) ties it to the phenomena and quantum information associated with this region.

The dimensionalities of $H$ and $B$ are denoted by $|H|$ and $|B|$, which are related to the area of the black hole's horizon measured in Planck units. The logarithms of these dimensionalities (i.e., $\log |H|$ and $\log|B|$) are proportional to the black hole's horizon area when we study the pure quantum state $\ket{\psi}$. Over time, the sizes of $|H|$ and $|B|$ decrease, illustrating the changes in the black hole dynamics as it ages. $H_ R$ starts restricted to a subset involved with the black hole dynamics, but the size of $H_ R$, denoted as $|H|$, increases over time, indicating that more states become relevant to the black hole's dynamics as time goes on. 

3) The third postulate involves the changes over time in the degrees of freedom or the "size" of $H_H$ and $H_B$. These changes in size are proportional to the area of the black hole's horizon, and as the black hole evolves, these values change. In other words, the behavior and properties of the black hole differ depending on whether $|R|$ is larger or smaller than the product $|H||B|$, giving rise to the distinction between "young" and "old" black holes. As a black hole ages, $|R|$ expands, altering the entanglement dynamics between $B$ and $H$ within the broader $H_{\text{outside}}$. $B$ and $H$ are highly entangled when a black hole is young. As the black hole ages, $B$ and $H$ become a small part of a much larger system described in Hilbert space. After a certain time, known as Page time, HH show that the combined system of $B$ and $H$ reaches a state that can be described using a simple mathematical expression involving the identity operator, which basically means the system has become uniform \cite{Harlow}. 

As time passes, the system reaches a state of maximal entanglement, where $BH$ (the combination of $B$ and $H$) is maximally entangled with $R$. 
HH decompose $H_R$ into a direct sum of subspaces, $H_{R_H}$ and $H_{R_B}$, and another space termed as $H_{\text{other}}$ (which does not contribute to the described entangled state): $H_R = (H_{R_H} \otimes H_{R_B}) \otimes H_{\text{other}}$. 
$H_{R_H} = |H|$ (or $H_{R_H} = |R_H|$) and $H_{R_B} = |R_B|$ are their dimensions which are set equal to $|H|$ and $|B|$ respectively for the entanglement to hold \cite{Harlow}. $R_B$ represents a portion of the surrounding space farther away from the black hole, where the Hawking radiation emitted from the vicinity of the black hole (including $B$) can be observed. $B$ and $R_B$ are maximally entangled, describing a high degree of scrambling. 

Let us keep in mind that we are still examining Charlie's perspective. 
Recall that the entire system (which encompasses both the black hole and the associated Hawking radiation, $H_R$) is represented by a pure quantum state, denoted $\ket\psi$, within $H_{\text{outside}}$. $\ket\psi$ is in a superposition of product states from the $H$, $B$, $R_H$, and $R_B$ Hilbert spaces:

\begin{equation} \label{eq1}
{\ket\psi} = \left(\frac{1}{\sqrt{|H|}} \sum_h \ket{h}_H \ket{h}_{R_H}\right) \otimes \left(\frac{1}{\sqrt{|B|}} \sum_b \ket{b}_B \ket{b}_{R_B}\right),
\end{equation} 
\vspace{1mm} 

\noindent where the terms $\ket{h}_H$ and $\ket{b}_B$ represent basis states in the spaces associated with the black hole and $B$, respectively, $\ket{h}_{R{_H}}$  and $\ket{b}_{R{_B}}$  represent basis states in the radiation Hilbert space corresponding to the states in the black hole and $B$.

Equation (\ref{eq1}) represents the state of an old black hole, written using Schmidt decomposition. It is factorized into two independent subsystems: one associated with $H$ and one associated with $B$. 
The fundamental argument of HH is grounded on equation (\ref{eq1}).
\vspace{1mm} 

The state in the equation is built such that each basis state in $H$ is entangled with a corresponding state in $R_H$ and similarly for $B$ and $R_B$. In other words, this structure reveals a deep and complex entanglement pattern: the states in $H$ are entangled with specific states in $R_H$, and those in $B$ are entangled with specific states in $R_B$. By setting up an entanglement between $H$ and $R_H$, and $B$ and $R_B$, we ensure the full state $\ket\psi$ is pure despite the subsystems $H$ and $B$ individually being in mixed states. 
\vspace{1mm} 

\emph{Now consider the perspective of Alice}, falling into a black hole, and compare her observations with those of Charlie, who is observing from outside the black hole. Here are the key points and steps of her argument: before reaching the singularity of the black hole, Alice also perceives her surroundings through a description based on a Hilbert space, $H_{\text{inside}}$, divided into several subspaces: $H_{\text{inside}}= H_A \otimes H_B \otimes H_R \otimes H_{H'}$,
where, $H_A$ represents the Hilbert space associated with $A$, which encompasses the field theory modes inside the black hole's event horizon from Alice's perspective. $A$ is just inside the event horizon of the black hole; $H_B$ and $H_R$ are outside the black hole and accessible to both Alice and Charlie; $H_{H'}$ is related to Alice's horizon, distinct from the black hole's horizon. From Alice's perspective, $H_H$ is absent. The region associated with $H_H$ is crossed by Alice in an extremely short time, meaning it has no operational significance to her; she wouldn't be able to conduct any meaningful measurements in such a short time frame. 

To avoid contradictions in Alice and Charlie's observations, an "overlap rule" ensures that both agree on the experiments' results they can communicate. Charlie and Alice, despite potentially having different perspectives, must come to the same conclusion regarding the density matrix of $H_B \otimes H_R$.

Alice would normally expect to see a smooth vacuum at the black hole horizon based on general relativity, with specific modes in $H_B$ and $H_A$ being closely entangled, reflecting a smooth vacuum at the horizon. However, this idea faces a significant problem due to the assumptions about what Charlie perceives at the black hole stage (specifically, being an old black hole).  

$A$, $B$, and $R$ would each have its associated Hilbert space representing all possible quantum states. $B$ is almost maximally entangled with $R_B$, which is perceivable by Charlie. Even though Alice cannot directly perceive $R_B$ (and Charlie cannot directly perceive $A$), the high degree of entanglement between $B$ and $R_B$ means that observing a part of the entangled system (i.e., $B$) gives information about the other part of the system ($R_B$). By the overlap rule, what Charlie perceives about the entanglement between $B$ and $R_B$ must also be true for Alice when she observes $B$.

However, because of the monogamy of entanglement, $B$ cannot be maximally entangled with $A$ and $R_B$ simultaneously. This contradicts the expectation of a smooth vacuum on the horizon for Alice. Recall from section \ref{1} that to resolve this contradiction, the AMPS argument posits that instead of a smooth vacuum, there is a firewall — a region of high-energy particles — at the horizon of an old black hole, which effectively annihilates anything or anyone, including Alice, that falls into it \cite{AMPS}; \cite{Harlow}. 

\subsection{Strong and standard complementarity}

To navigate this problem, one can choose two potential directions \cite{Harlow}:

1) \emph{Strong complementarity}: HH have defined Alice’s and Charlie’s theoretical frameworks. Alice's theory doesn't directly relate to Charlie's in this approach. They only need to agree on the experimental results visible to both. 

The event horizon serves as a barrier, making direct access to information in $R_B$ inherently impossible from Alice's localized perspective near the black hole. The states in $R_B$ are maximally entangled with those in $B$, i.e., knowing the state in $B$ allows one to predict the state in $R_B$ perfectly, and vice versa. Over time, the information about the black hole gets scrambled and distributed over a very wide area, including $R_B$. In Charlie’s theory, $B$ and $R_B$ are significantly entangled due to the dynamics of black hole evaporation. This entanglement ties up $B$, preventing it from being highly entangled with $A$. But Alice cannot access quantum states in the Hilbert space $H_{R_B}$ associated with $R_B$. So, in Alice's theory, there are two potential solutions to address this problem: 

1. Disentangle $H_{R_B}$ from $H_B$: This would involve creating a theory where the states in $R_B$ are no longer entangled with those in $B$. This freeing up of $B$ would entangle it with $A$.

2. Remove $H_{R_B}$ from her theory entirely. Alice’s theory would ignore the states in $R_B$, treating them as irrelevant. This would again free up $B$ to become entangled with $A$. This would grant Alice a smooth journey across the horizon.

Thus, according to strong complementarity in Alice's frame, certain manipulations or neglect of $H_{R_B}$ are postulated to allow a self-consistent description of a smooth horizon, harmonizing the dual descriptions by Charlie and Alice. Strong complementarity basically advocates for a kind of duality in the descriptions, where both are valid in their respective domains. Still, neither description can be globally valid, thereby complementing each other.

2) \emph{Standard complementarity}: While described using different theoretical frameworks, Alice's and Charlie's observations can be compatible because Alice's theory can be seen as a subset embedded in Charlie's more encompassing theory. A theoretical framework is developed where Alice's interior operators to describe phenomena in $H_A$ are formulated using the exterior operators that Charlie would use to describe $H_{R_B}$. This attempt is made to avert the problem of firewalls and maintain the notion of a smooth space at the horizon, referring to it as $A=R_B$ \cite{Bousso}. 

This equality symbolizes an effort to harmonize the two perspectives (Alice’s and Charlie's) by identifying the region described by Alice ($A$) with $R_B$ in Charlie's description. By developing a theory where Alice's interior description (on $H_A$) is formulated using exterior operators that Charlie would use to describe $H_{R_B}$, standard complementarity aims to avoid the problem of firewalls and uphold the idea of a smooth transition at the event horizon \cite{Susskind1}; \cite{Susskind2}.

However, identifying Alice’s $A$ with Charlie’s $R_B$ introduces a notable problem. It doesn't restrict Alice from making direct measurements in $R_B$, which, according to the broader theory (Charlie's perspective), she shouldn't be able to access directly. This results in a paradox where Alice, while theoretically unable to access $R_B$, finds herself able to do so according to this approach, thereby contravening the theory that posits complementarity.

\subsection{The unitary transformation and its inverse}

In what follows, I show that HH introduce a new perspective that involves computational complexity to help address the issues raised by the AMPS paradox. They bring a different angle to this discussion, focusing on the computational complexity involved in reconstructing the interior of a black hole (Alice's point of view) from the information outside of it (Charlie's point of view). They argue that the task of Charlie reconstructing what is inside (and thus detecting a firewall) would involve an intractable computational task. I demonstrate that HH bring a fresh perspective that adds a nuanced layer to the standard complementarity, but they navigate around the standard complementarity rather than extend it.

Initially, HH phrase the discussion in terms of Charlie's Hilbert space. Since both Charlie and Alice are required to agree on the density matrix that describes the system in terms of $H_B$ and $H_R$, it implies that a consistent representation of $\ket{\psi}$ can be achieved from the perspective of Charlie's Hilbert space. Hence, despite potentially different vantage points, Alice and Charlie will converge on a shared description, abiding by the constraints of the overlap rule.  

First, HH specify that in the Schmidt decomposition, one describes the state of the old black hole in terms of a basis that effectively separates the early and late-time radiations, see equation (\ref{eq1}). However, HH will use a particular basis (denoted as "computational basis") to describe the radiation field $H_R$, which will be convenient for Alice to work with. They write an equation that denotes the specific computational basis state ($\ket{bhr}_R$) in $H_R$ using $n$, $k$, and $m$ that describe various aspects of that state. Working exclusively with this basis is intended to simplify calculations for Alice as she tries to work through the decoding problem with a basis where the mathematical expressions become more tractable. 
Alice uses $n$, $k$, and $m$ to represent different quantities related to her problem of decoding information from the black hole's Hawking radiation: $n$ represents the total number of qubits involved in the problem. This is related to the logarithm to base $2$ of the dimension of $R$: $n\equiv \log_2|R|$; $k$ represents the number of qubits associated with $H_B$; and $m$ represents the number of qubits associated with $H_H$. We can think of $k+m$ as the number of qubits remaining in the black hole. $H_H$, along with $H_B$, forms the entirety of the black hole's state that Alice is interested in for her problem. 

Second, HH define $U_R$ as a particular (scrambling) unitary transformation on $H_R$ and write $\ket{\psi}$ [equation (\ref{eq1})] in the computational basis: 

\begin{equation} \label{eq2}
{\ket\psi} = \frac{1}{\sqrt{|B||H|}} \sum_{b,h} \ket{b}_B \ket{h}_{H}U_R\ket{bh0}_{R}.
\end{equation}

\noindent $U_R$'s exact form or structure depends on the details of the black hole's initial state and the quantum gravity, a yet-to-be-realized theory. The discussion here pertains to Schwarzschild black holes instead of AdS eternal black holes that do not evaporate and maintain a dynamic equilibrium with Hawking radiation. This dynamic equilibrium significantly changes the conditions and assumptions on which the HH decoding task and the $U_R$ are defined. HH note that "Big AdS black holes do not evaporate at all, so the AMPS argument does not directly apply to them, but arguments have been put forward suggesting that they nonetheless have firewalls" \cite{Harlow}.

Alice's challenge is applying the inverse of $U_R$, denoted $U^{\dagger}_R$, to the Hawking radiation. This allows her to confirm the entanglement between $H_B$ and $H_{R_B}$. Applying $U^{\dagger}_R$ reverses the transformation effected by $U_R$, decoding the information encoded in the Hawking radiation. The goal is to retrieve the information encoded in the entangled states between $H_B$ and $H_{R_B}$ from the Hawking radiation. Through decoding, Alice can confirm the entanglement between $H_B$ and $H_{R_B}$, gaining insight into the states and dynamics occurring in $R_B$ through her operations on the radiation emitted by the black hole without directly accessing $R_B$. By utilizing the decoded information from the Hawking radiation, Alice can avoid the need for direct measurement in $R_B$, thus avoiding the paradox arising from her ability to access information in $R_B$ directly according to the standard complementarity approach. In other words, Alice can bypass the complications brought about by the firewall argument while upholding the principles inherent in the black hole complementarity.

However, the time Alice would need to decode the information is exponential and proportional to $2^{k+m+n}$. This time complexity is extremely large, indicating a computational task of astronomical proportions. Since the black hole is old, it implies that $n$ (representing the age parameter) is significantly larger than the sum of $k$ and $m$. This condition allows us to derive a lower bound for the reduced time complexity expression: $n>k+m$. Using this condition in $2^{k+m+n}$, we can establish a lower bound on the reduced time complexity as at least $2^{2{(k+m)}}$. But this expression also involves an exponential function and, therefore, would be extremely large for non-trivial values of $k$ and $m$. 

Thus, reversing $U_R$ exactly would be practically impossible for Alice. Hence, a framework is required to quantify how close Alice can reach the perfect task. HH use a concept of "trace norm," which allows them to define a notion of "closeness," how close Alice needs to get to test the entanglement accurately \cite{Harlow}. In other words, they aim to find a unitary operation close enough to the ideal operation $U^{\dagger}_R$ such that the error in the subsequent measurements is within acceptable bounds. Defining a threshold for the closeness using trace norm would define how accurately Alice needs to perform the unitary transformation to test the entanglement reliably. It establishes a criterion for the allowable error in Alice's operation, such that she can still validate the entanglement through her measurements. By working within this framework, Alice aims to construct a unitary transformation that, while not the same as the ideal $U^{\dagger}_R$, is close enough according to this defined metric to allow her to verify the entanglement adequately. This approach acknowledges the practical difficulties and uncertainties in constructing the exact unitary transformation and provides a pathway to achieve Alice’s goal within tolerable error margins. 

By allowing for a non-ideal transformation, where Alice tries to get close enough through trace norms, there might be a reduction in the complexity of the task at hand. However, even when settling for a non-ideal transformation, Alice still faces an enormously complex computational task. Though the requirement is somewhat relaxed compared to achieving the ideal transformation, the task remains a high-order polynomial time problem, implying a computational duration that vastly exceeds the black hole's lifespan. 

\section{HH provide Alice with a quantum computer} \label{3}

\subsection{Alice is distilling the radiation with a quantum computer}

Alice is on a mission to decode the entanglement between $R$ and $B$, aiming to transform it into a format that is easier to analyze. To accomplish this, Alice uses a quantum computer whose initial state is defined in a new Hilbert space $H_C$. This computer interacts with the radiation Hilbert space $H_R$, undergoing a unitary evolution described by $U_{\text{comp}}$. In other words, $U_{\text{comp}}$ operates on a larger Hilbert space that includes $H_R$ and $H_C$: $H_R \otimes H_C$. Alice employs $U_{\text{comp}}$ (similar to $U^{\dagger}_R$) to reverse the effects implemented by $U_R$ to distill $R_B$. This intends to decode the information entangled with $B$ during the evolution governed by $U_R$. 

However, Alice's major challenge is finding the appropriate initial state for her computer:

\begin{equation}
U_{\text{comp}}: U_R |bh0\rangle_R \otimes |\psi\rangle_C \rightarrow |\text{something}\rangle \otimes |b\rangle_{\text{mem}}.    
\end{equation}
\vspace{1mm} 

\noindent Alice is looking for a particular initial state for her computer, denoted $\ket{\psi}_C$, such that under $U_{\text{comp}}$, this state interacts with the states from the radiation field $U_R\ket{bh0}_R$ to produce a final state where the qubits representing $\ket{b}$ (which are associated with the $B$ basis) are stored in the first $k$ qubits of the memory of her computer, separated from the other parts of the system described by $\ket{\text{something}}$. In other words, Alice is trying to isolate information about $B$ into her computer’s memory \cite{Harlow}.

However, finding the initial state $\ket{\psi}_C$ for the computer to start the computation is extremely challenging. Even if Alice finds one, the probability that it successfully facilitates the computation is minuscule, being exponentially small in terms of the dimensions of the Hilbert space. To estimate the likelihood of finding a successful initial state, HH use the trace norm to create a set close to which every pure state in the Hilbert space can be found. Even after considering various potential unitary evolutions, the probability of finding a successful initial state remains extraordinarily small. The time required to find a suitable $U_{\text{comp}}$ by random chance is identified as the quantum recurrence time, the time scale over which a quantum system revisits a particular state it was in at some earlier time, due to its natural dynamics. The recurrence time is immensely long, reaching up to $10^{10^{40}}$ years. The recurrence time in which a quantum system revisits near an initial state by pure chance is especially long for the complex system Alice is dealing with, involving a black hole and Hawking radiation. This time scale is doubly exponential in terms of the entropy of the whole system, i.e., it increases extremely fast as the system's complexity increases. Thus, the recurrence time is potentially longer than the lifespan of a Schwarzschild or an astrophysical black hole.

Despite the immense challenge posed by the quantum recurrence time, Alice has another strategy up her sleeve. Instead of the brute force approach, waiting for the right $U_{\text{comp}}$ to occur by chance, she aims to actively find it by exploiting patterns in how $U_{\text{comp}}$ evolves. If there are any predictable structures in its evolution, she can use this information to guide her search, significantly reducing the time it would take to find the right $U_{\text{comp}}$. By leveraging the structures in $U_{\text{comp}}$'s evolution, Alice can potentially reduce the computational time from a double exponential dependency on the entropy of the radiation to a single exponential dependency. This is a very long time since single exponential growth is very fast. But it is much more manageable compared to a double exponential growth. So, this strategy offers a glimmer of hope; despite the initially grim prognosis suggested by the quantum recurrence time, if Alice can understand and leverage the underlying physics and mathematical properties of her system well enough, she might be able to achieve her goal in a "reasonable" amount of time, though still astronomically long from a human perspective \cite{Harlow}.

\subsection{The Solovay-Kitaev theorem}

HH then ask \cite{Harlow}: how many quantum gates would be necessary to implement a unitary transformation as complicated as $U_R$? They show that the number of gates required scales with a single exponential function of the number of qubits ($n$), $2^{2n}$ times a logarithm of a parameter ($\epsilon$): $2^{2n}\log(\frac{1}{\epsilon})$.  

This is a significant result because it means the computational time is a lot less than previously assumed based on a crude double exponential scaling (for instance, $2^{2^{n}}$). This improvement is partly credited to the Solovay-Kitaev theorem, which suggests that an efficient sequence of gates can be found to perform any unitary operation to a high degree of precision. Despite this improvement, reducing the computational time further seems unlikely, indicating that $U_R$ and $U^{\dagger}_R$ are still highly non-trivial tasks even under optimal conditions. Adjusting the model, such as using different gates or more complex quantum entities like qutrits (which have three basis states) instead of qubits, doesn't fundamentally change the $2^{2n}$ scaling of the problem. 

We can speed up computations significantly compared to initial expectations. However, we are still looking at a process that requires a time that scales exponentially with the number of qubits, which indicates extremely long computation times for complex operations involving many qubits. This makes Alice's task of implementing $U_R$ within a reasonable time frame highly unlikely.

\section{HH combine black hole physics and computing} \label{4}

\subsection{Taking into consideration the dynamics of the black hole} 

HH then suggest that the dynamics of a black hole constrain $U_R$ in a way that could help Alice implement it faster. A transformation $U_{\text{dyn}}$ is defined as a unitary transformation that operates on different microstates of a black hole, and “it seems quite reasonable to assume that $U_{\text{dyn}}$ can be generated by a polynomial number of gates,” i.e., a computational operation that doesn't require an astronomical number of quantum gates to perform \cite{Harlow}. I am quoting this phrase because I find it problematic from a conceptual and philosophical standpoint. I elaborate on this perspective below. 

HH define the state $\ket{\psi}$ as arising from the action of the polynomial
size circuit $U_{\text{dyn}}$: 

\begin{equation}
U_{\text{dyn}}{\ket0} _{BHR}= \frac{1}{\sqrt{|B||H|}} \sum_{bh} \ket{b}_B \ket{h}_{H}U_R\ket{bh0}_{R}.
\end{equation}

\noindent They provide a more comprehensive explanation of this matter. Alice wants to determine if a small circuit for $U_{\text{dyn}}$ implies a small circuit for $U_R$. If such a circuit exists, she could more easily decode $R_B$ from the Hawking radiation. 
The matrix $U_R$  is derived from $U_{\text{dyn}}$, which, unlike $U_{\text{dyn}}$, is sensitive to the system's initial state. For the sake of simplicity, an initial state is chosen with all bits set to zero. HH write an expression for $U_{\text{dyn}}$ acting on this initial state, intending to learn more about $U_R$:
 
\begin{equation}
U_{\text{dyn}}\ket{00000}_\text{{init}}  \approx \frac{1}{\sqrt{|B||H|}} \sum_{b,h} \ket{b}_B \ket{h}_{H}U_R\ket{bh0}_{R}.    
\end{equation}

Some physical assumptions about $U_{\text{dyn}}$ are necessary to further this exploration. While the precise dynamics of quantum gravity remain undefined, HH refer to theories like AdS/CFT and matrix theory, which give ground to assume that $U_{\text{dyn}}$ can be generated through polynomial numbers of gates in a small circuit \cite{Harlow}. So, the core search then revolves around whether there is a small circuit for $U_{\text{dyn}}$ that ensures a small circuit for $U_R$. This would imply that if affirmed, Alice could feasibly decode $R_B$ from the Hawking radiation.
However, it seems like there is a challenge here, citing the eternal nature of black holes as posited by AdS/CFT theory, which ostensibly disputes the presence of a firewall paradox. This presents a theoretical contradiction, introducing complexity and potential disagreement in integrating these theories with the problem at hand. 

Next, disregarding the complication, HH proceed to articulate their argument: $U_{\text{dyn}}$ can be expressed as a product of $U_R$ and another operation called $U_{\text{mix}}$. $U_{\text{mix}}$ is a unitary operation that creates entanglement between different subsystems, mixing or scrambling the information within them to create a highly entangled state.

Here, $U_{\text{mix}}$  is conceptualized as a straightforward circuit that entangles the initial four subfactors in the chosen initial state. HH emphasize the simplicity of implementing $U_{\text{mix}}$, noting that a universal circuit would suffice:

\begin{equation}
U_{\text{mix}}\ket{00000}_\text{{init}} = \frac{1}{\sqrt{|B||H|}} \sum_{b,h} \ket{b}_B \ket{h}_{H}U_R\ket{bh0}_{R}.    
\end{equation}

HH then define a new operator \noindent $\tilde{U}_R$ as: $\tilde{U}_R=U_{\text{dyn}}U^{\dagger}_{\text{mix}}$, which has the property:

\begin{equation}
\tilde{U}_R \frac{1}{\sqrt{|B||H|}} \sum_{b,h} \ket{b}_B \ket{h}_{H} \ket{bh0}_{R}= \frac{1}{\sqrt{|B||H|}} \sum_{b,h} \ket{b}_B \ket{h}_{H}U_R\ket{bh0}_{R}.    
\end{equation}

HH argue that a small circuit can implement $\tilde{U}_R$. It undoes the effects of $U_R$ when applied to a specific superposed state, a complex combination of states from $B$ and $H$ \cite{Harlow}. 

\subsection{Alice again confronts complications}

The above process is highly non-trivial because, as seen in the previous section, the new inverse operator $\tilde{U}_R$ is intricately dependent on the precise initial states of the qubits in $B$ and $H$, which are part of an entangled system associated with a black hole. Applying $\tilde{U}_R$ ideally would unscramble the quantum information, extracting it from the entangled state and converting it into a format that can be more straightforwardly accessed and analyzed.
Although it seems that $\tilde{U}_R$ could be utilized to decode the information, it encounters a pivotal issue; the operation of $\tilde{U}_R$ involves all qubits, including those in $B$ and $H$, to which Alice doesn't have access. Hence, implementing $\tilde{U}_R$ as is would be infeasible. 
Alice, therefore, considers a strategy where she replaces the unavailable qubits from $B$ and $H$ with some ancillary qubits in a random state, hoping this would allow her to use $\tilde{U}_R$ to undo $U_R$. The difficulty remains as $\tilde{U}_R$ is fundamentally tied to the initial states of the $B$ and $H$. 

Ultimately, it is highly unlikely for Alice to find a simple, small circuit to implement $U_R$ due to the constraints imposed by $U_{\text{dyn}}$ and the complexity arising from $U^{\dagger}_R$. Because Alice doesn't have access to all the qubits she needs, she can't directly apply $U^{\dagger}_R$ to reverse the dynamics and extract the information encoded in $R_B$. The transformations she would like to apply, including $\tilde{U}_R$, are sensitive to the states of all the zones, including $B$ and $H$, which she can't access. Therefore, Alice can't take a straightforward approach to reverse the dynamics in a manageable amount of time, i.e., in a time that is a polynomial entropy function. So, Alice is stuck with having to try a brute force strategy, where she uses a tremendous number of operations ($2^{n+k+m}$ gates, which indicates a huge, potentially impractical number) to construct $U^{\dagger}_R$. HH, therefore, suggest a kind of pessimism or skepticism that there would be an easy solution to Alice's problem \cite{Harlow}.

\section{HH transform a physical problem into a quantum coding problem} {\label 5}

\subsection{Alice is utilizing an error-correcting code}

HH then hint toward the possibility of exploring the problem further through the lens of error-correcting codes and complexity theory, albeit acknowledging the sheer challenge posed by the exponential increase in complexity \cite{Harlow}. They recast Alice’s task as a quantum coding problem. 

Alice's main difficulty is retrieving information, especially when dealing with errors introduced by the black hole environment, which causes erasures in $B$ and $H$ (which define the state of the black hole). Recall that Alice has no access to $B$ and $H$. This lack of access is represented mathematically as erasures in these systems. These erasures pose a significant challenge to Alice's task, as they take away information that is vital for her to be able to successfully decode the information about the black hole's initial state. Alice uses the error-correcting code to recover information that has been erased and affected by the interaction of the black hole with its environment.

HH write equation (\ref{eq2}) in an equivalent way:

\begin{equation} \label{eq3}
\begin{aligned}
{\ket\psi} = \frac{1}{\sqrt{|B|}} \sum\ket{b}_B \ket{\overline{b}},
\ket{\overline{b}} \equiv \frac{1}{\sqrt{|H|}} \ket{h}_{H}U_R\ket{bh0}_{R}.
\end{aligned}
\end{equation}

\noindent The state $\ket{\psi}$ is expressed as a superposition of $\ket{b}_B$ and $\ket{\overline{b}}$, where $\ket{\overline{b}}$ is defined in terms of $\ket{b}_B$ and $U_R$. HH are trying to simplify the problem by focusing on a smaller Hilbert space, where $\ket{\overline{b}}$ is a basis for a $k$ dimensional subspace of $H_H \otimes H_R$. But they also underscore the intertwined nature of the state $\ket{\psi}$ in the black hole and the Hawking radiation system, with its behavior being determined by interactions between multiple different Hilbert spaces. 

Alice creates a quantum code using $U_{\text{enc}}$, which is built from $U_R$ and $U_{\text{mix,H}}$: $U_{\text{enc}}\equiv U_RU_{\text{mix,H}}$. 
$U_{\text{mix,H}}$ is a simple entangling transformation analogous to $U_{\text{mix}}$, which is applied to a subset of the qubits representing a part of the radiation emitted. It affects the $m$ qubits of $H$ and the $n + 1$ to $n + m^{th}$ qubits of $R$. This encoded information involves entangling pairs of qubits from $H$ and $R$ to produce new states. 
Up to this point in their discussion, HH have focused on the errors that arise from the fact that Alice does not have direct access to all the necessary information from $B$ and $H$ to decode the information she seeks straightforwardly. However, they point out that this isn't Alice's task's only source of potential errors. They bring attention to another substantial challenge. For instance, the black hole emits Hawking radiation, including hard-to-detect gravitons, let alone coherently manipulate.  Thus, information about the black hole's state is carried away with that radiation, and some of it is lost.

When Alice applies $U_{\text{enc}}$, she identifies a subset of states (subspace) of the total Hilbert space within the larger Hilbert space. By narrowing it down to a specific subspace, Alice is reducing the computational complexity and resource requirements of the task at hand. Operating in a smaller subspace allows a more streamlined approach to information retrieval, bypassing the need to deal with a prohibitively large set of all possible quantum states in the full Hilbert space. This subspace effectively helps in encoding the information about the black hole state and aids in protecting and retrieving the information even after some amount of information loss and erasure. Moreover, the universe of all possible quantum states in the Hilbert space is vast, and working with it directly would be computationally prohibitive. By identifying a subspace through the $U_{\text{enc}}$ transformation where the black hole information is encoded, Alice can work with a reduced, more manageable set of states, facilitating the decoding process.

When Alice wants to retrieve the information, she is looking to decode the information stored in this subspace. This decoding process reverses $U_{\text{enc}}$ to retrieve the original information, or at least the pertinent part of it, even after parts of the system have been erased. However, reversing $U_{\text{enc}}$ to decode the information isn't straightforward. Alice uses a set of quantum operations, a universal gate (CNOT gates and Hadamard transformations), to develop $U_{\text{enc}}$, and she tries to identify a suitable decoding process that can retrieve the original information from the encoded state. To do this, Alice introduces an additional system denoted as $B'$, with the same number of qubits as $B$, and forms a new, more extensive set of encodings. The crucial part of the decoding process involves a complex unscrambling of the entanglements between $B'$ and $R$ through another set of transformations $U_{\text{enc}}$, which is constructed from $U_{\text{dyn}}$  and the transformation that involves the new system $B'$, $U_{\text{mix, B'}}$.  

More specifically, HH want to establish $U_\text{dyn}$ as the encoding transformation $U_{\text{enc}}$ and $U_R$ as a correction operation. However, there is an issue with the dimensionality of the code space, which they address by introducing an additional system $B'$ with the same number of qubits as $B$. They then apply a transformation $U_{\text{mix},B'}$ to entangle $B'$ with $B$, thereby creating a $k$-qubit code subspace within a larger $2k+m+n$ qubit Hilbert space using $U_{\text{enc}}$. $U_{\text{enc}}$ is defined by applying $U_{\text{mix},B'}$ followed by $U_\text{dyn}$. $U_{\text{enc}}$ operates on initial states to generate a new state that incorporates $B$, $B'$, $H$, and $R$, such that: 

\begin{equation}
U_\text{enc}\ket{b'}_{B'} \ket{0}_{BHR}=\frac{1}{\sqrt{|B||H|}} \sum_{bh} U_{\text{mix},B'} \ket{b'b}_{B'B} \ket{h}_{H}U_R\ket{bh0}_{R}.
\end{equation}

HH acknowledge the presence of errors introduced by the environment, conceptualized as erasures affecting both $B$ and $H$. To restore the initial state by unscrambling the entanglement between $B'$ and $R$, one needs to implement $U^{\dagger}_R$. By applying $U^{\dagger}_R$ on the newly defined state, and subsequently acting with $U_{\text{mix},B'}$ , the information $b'$ is recovered. This process involves replacing the second element of each pair of qubits from the first $k$ qubits of $R$ instead of from $B$, which isn't accessible anymore. 
This produces a state:

\begin{equation}
\ket{b'}\frac{1}{\sqrt{|B||H|}} \sum_{bh} \ket{b}_B \ket{h}_{H}\ket{bh0}_{R},    
\end{equation}

Following the recovery of $b'$, the initial state can be restored in polynomial time using ancillary qubits to reset $BHR$ to the state $\Ket{0}_{BHR}$ and then utilizing $U_{\text{enc}}$ to revert to the desirable state formed by the encoding transformation involving $B'$ and $BHR$ \cite{Harlow}. 

By introducing a new system $B'$, Alice adds a new set of variables. $B'$ interacts with the radiation ($R$); a new set of entangled states between $B'$ and $R$ are formed through this interaction. These entangled states have information about the black hole. $U_{\text{enc}}$ is a set of operations designed to unscramble the entangled states formed by interactions between $B'$ and $R$. It uses the dynamics of the black hole and the transformations involving $B'$ to create a scenario where the entangled information can be unscrambled, running some processes in reverse to recover the original data. Alice uses $U_{\text{enc}}$ to recover the original data encoded in the entangled states of the black hole information. This is trying to reverse the effects of the black hole dynamics on the information by decoding the previously encoded information in the scrambled, entangled states. By utilizing a more comprehensive set of encodings involving the additional system $B'$ and implementing a complex transformation to unscramble the entanglements, Alice aims to recover the original data and thereby solve the problem of decoding the information about the state of the black hole, circumventing the issues encountered with not having access to all the required qubits in the previous setup. This strategy is grounded on the hope that by properly configuring the new system $B'$ and skillfully using $U_{\text{enc}}$, Alice can create a pathway through the entanglement structure that allows her to access the information she is after, even without direct access to all parts of the system. 

\subsection{A beacon of hope amidst lingering challenges}

For the channel just constructed, $n$ is the total number of bits that can theoretically be lost due to various errors or complications while still retaining the ability to correct those errors and successfully decode the information. $2k + m + n$ breaks down this $n$ into components corresponding to different aspects or parameters of the black hole model. So, $k+m$ bits have already been unavoidably lost because Alice doesn't have access to $B$ and $H$. $\frac{n - k - m}{2}$ is the remaining number of bits Alice can afford to lose due to other errors and still maintain the possibility of successful error correction. $\alpha$ represents the fraction of the radiation that consists of gravitons. 
HH then make a critical point: even if all the gravitons (which constitute less than half of the radiation) are lost, Alice could theoretically still extract the necessary entanglement information accurately because she still has room for error correction — she can still lose up to $\frac{n - k - m}{2}$ bits and remain within the bounds for successful error correction. Even if Alice cannot control and measure the gravitons, she can wait until enough radiation has been emitted such that the information about the interior of the black hole becomes accessible in the radiation. She can still successfully decode the necessary information to solve the task.

However, HH note two important complications \cite{Harlow} 

1) Computational complexity comes into play here as a significant roadblock. Extracting the information from the radiation is an immensely complex computational task. This complexity grows with the size of the black hole and the amount of information to be retrieved. 

2) The task must be done before the black hole fully evaporates, setting a strict time limit on the procedure. Thus, while theoretically possible, practically carrying out the AMPS experiment to solve the black hole information problem without considering computational complexity would be extraordinarily challenging and likely unfeasible due to the computational resources required. Thus, Alice might not have enough time to complete her task before the black hole evaporates, hence not being able to rule out the presence of a firewall definitively.

\section{A physics problem becomes a quantum computing problem} \label{6}

\subsection{HH's \textup{\textsf{Error Correctability}} problem}

HH introduce a computational problem termed \textup{\textsf{Error Correctability}} \cite{Harlow}. This problem arises in the context of $B$, $H$, and $R$. The state of this system is transformed by $U_{\text{dyn}}$ acting on an initial state, resulting in a new state $\ket{\psi}_{BHR}$, which exhibits maximal entanglement between $B$ and $HR$.

\begin{equation}
\ket{\psi}_{BHR} = U_{\text{dyn}}\ket{000}_{BHR},   
\end{equation}

The crux of the \textup{\textsf{Error Correctability}} problem is to ascertain whether it is possible to decode the maximal entanglement with $B$ by only using information from $R$. This problem has a \textup{\textsf{QSZK}} proof, which is characterized by a verifier checking the results after a prover implements a quantum error correction operation on $R$, despite the potential for this process to take an exponentially long time. 

HH replicate the kind of maximal entanglement generated during the dynamical evolution governed by $U_{\text{dyn}}$ using a different strategy involving a noise operator $U_{\text{noise}}$. In the strategy: 

\begin{equation}
\ket{\psi}_{BHR} =U_{\text{noise}}\ket{000}_{BHR},    
\end{equation}

\noindent is utilized to create a state where the $BH$ and $R$ systems become maximally entangled. By applying $U_{\text{noise}}$ repeatedly (i.e., working with $k$ copies of $\ket{\psi}$, noted as $\ket{\psi}^{\otimes k}$) one converges towards a state with near-maximal entropy concentrated in a typical subspace of the tensor product of $k$ copies of the $B$ and $H$ Hilbert spaces.
However, HH mention that this approach, while similar to the true maximal entanglement generated by $U_{\text{dyn}}$, might be slightly weaker. This suggests that while this strategy can approach the maximal entanglement that $U_{\text{dyn}}$ would generate, it might not fully replicate it, potentially missing some correlations or not fully reaching the maximal entropy that would signify truly maximal entanglement. Hence, although this strategy involving $U_{\text{noise}}$ can yield a state with a high degree of entanglement between the $BH$ and $R$ subsystems, it seems it might not exactly reproduce $U_{\text{dyn}}$.

\subsection{Evaluating Alice's task through the lens of \textup{\textsf{QSZK}} and \textup{\textsf{BQP}} complexity classes}

HH refer to the following complexity classes \cite{Harlow}: the \textup{\textsf{BQP}} complexity class, which contains decision problems that can be solved by a quantum computer in polynomial time, with an allowable error probability. 
The \textup{\textsf{SZK}} complexity class: a class of problems where solutions can be verified using classical computers, and where the "yes" instances can be proven with zero-knowledge proofs, where the verifier learns nothing other than the fact that the statement is true.
\textup{\textsf{QSZK}} (Quantum Statistical Zero Knowledge), a quantum analog of the \textup{\textsf{SZK}} complexity class. This complexity class contains \textup{\textsf{BQP}}. \textup{\textsf{QSZK}} is a complexity class that refers to the set of computational problems with yes/no answers where a (potentially dishonest) prover can always convince a verifier of the yes instances but will fail with high probability to convince them of the no instances using quantum computations.

HH identify the \textup{\textsf{Error Correctability}} problem as a \textup{\textsf{QSZK}}-complete problem \cite{Harlow}. A problem that is \textup{\textsf{QSZK}}-complete represents the hardest problems in the \textup{\textsf{QSZK}} class. If there is a polynomial time solution for one \textup{\textsf{QSZK}}-complete problem, then there are polynomial time solutions for all problems in the \textup{\textsf{QSZK}} class. In other words, if \textup{\textsf{Error Correctability}} is identified as a \textup{\textsf{QSZK}}-complete, then it has two properties:

1) It is in \textup{\textsf{QSZK}} and can be solved using a quantum computer with the help of an all-powerful, though potentially dishonest, prover.

2) Any problem in \textup{\textsf{QSZK}} can be reduced to the \textup{\textsf{Error Correctability}} problem through a polynomial-time algorithm. 

Solving the \textup{\textsf{Error Correctability}} problem would imply a solution for every problem in the \textup{\textsf{QSZK}} class.

HH argue that if \textup{\textsf{QSZK}} were equal to \textup{\textsf{BQP}}, then all the problems that can be solved with a quantum verifier and a prover (\textup{\textsf{QSZK}}) can also be solved with just a quantum computer operating in polynomial time (\textup{\textsf{BQP}}). In other words, the prover does not add any computational power; a quantum computer alone is sufficient to solve these problems. If this decoding can be done on a quantum computer in polynomial time, it would mean that all \textup{\textsf{QSZK}} problems can be solved in \textup{\textsf{BQP}}, implying \textup{\textsf{QSZK}} = \textup{\textsf{BQP}} \cite{Harlow}.
HH are asking whether the decoding process required to solve the \textup{\textsf{Error Correctability}} problem can be done in polynomial time with the size of the circuit used to represent correctable noise denoted by $U_{\text{dyn}}$. They reiterate that finding a polynomial-time solution to the decoding problem (concerning the size of the $U_{\text{dyn}}$) would mean that the \textup{\textsf{Error Correctability}} problem (and hence all problems in \textup{\textsf{QSZK}}) can be solved in \textup{\textsf{BQP}} time. This would establish that \textup{\textsf{QSZK}} = \textup{\textsf{BQP}}, showing that a quantum computer, without the help of a prover, can solve all problems in \textup{\textsf{QSZK}}. 

HH conclude the discussion by providing the following evaluation \cite{Harlow}: 

\begin{quote}
[…] by recasting Alice's problem as quantum error
correction, we have set it into a framework where there are general arguments that such problems likely take exponential time to solve. Moreover the practical difficulties of doing the experiment, in particular the problems associated with measuring gravitons, further increase the difficulty of this computational task. We did not quite manage to prove that her task is \textup{\textsf{NP}}-hard at fixed $k$, but it is almost certainly at least \textup{\textsf{QSZK}}-hard and there are strong reasons to believe that such problems can't be solved in polynomial time on a quantum computer. From the computer science point of view, it would be extremely surprising if implementing $U^{\dagger}_R$ did not require exponential time.     
\end{quote}

\subsection{Aaronson's improvements to the complexity assumptions underlying HH's task}

Scott Aaronson suggested improvements to HH's decoding task \cite{Aaronson}. He proposed improvements to the complexity assumptions underlying HH's decoding task. Instead of working directly within the framework of complexity theory classes, he ground the problem's difficulty in established cryptographic concepts, leveraging the one-way functions (OWF)s and introducing hardcore predicates to argue the hardness of the decoding task \cite{Aaronson}.

Aaronson endeavors to demonstrate that if there are injective OWFs, functions that are easy to compute in the forward direction but hard to invert even by quantum computers, then the HH decoding task is indeed hard. This approach aims to ground the decoding task in more widely accepted theoretical foundations, 
thereby enhancing the robustness of the argument concerning the inherent difficulty of the HH decoding task by rooting it in established cryptographic principles based on OWFs. Information is encoded using these OWFs. To decode information from a portion of these states (i.e., from the Hawking radiation part), one needs to find the inverse of the OWF on the encoded information – which is assumed to be computationally hard because of the property of OWFs. Therefore, if someone could solve the HH decoding task, it would mean they have a method to invert OWFs. This has far-reaching implications, including breaking the widely accepted cryptographic systems that rely on the hardness of inverting OWFs.

Aaronson further argues that solving the HH decoding problem would require at least as much computational resource as the hardest problems in the collision problem class. He connects the ability to invert OWFs and the ability to find collisions. If one could efficiently invert an OWF, one could find collisions, undermining the security assumptions of cryptographic systems based on these functions. Aaronson's statement emphasizes the fundamental difficulty in decoding the Hawking radiation as it would imply being able to solve other hard problems in computer science and cryptography. 

Aaronson is tying the hardness of a physics problem – decoding Hawking radiation and the firewall paradox – to established hard problems in computer science, grounding the task's believed difficulty in well-established complexity theory \cite{Aaronson}. Through his improvements to the HH decoding task, Aaronson brought a deeper computer science perspective into the discussion surrounding the firewall paradox, specifically leveraging concepts from the computational complexity theory and cryptography. By framing the HH decoding task as a problem in computer science and showing that it's at least as hard as inverting one-way functions, Aaronson translated a physics problem into the language and framework of computer science. This approach not only allows for a more robust argument regarding the complexity of the problem but also facilitates interdisciplinary dialogue and understanding by linking the physical scenario to well-established concepts in the computer science domain.

\vspace{2mm} 

\section*{Acknowledgement}

\noindent This work is supported by ERC advanced grant number 834735.

\end{document}